\documentclass[letter,twocolumn]{jpsj3}

\usepackage{graphicx}
\usepackage{dcolumn}
\usepackage{bm}
\usepackage{color}
\usepackage{amsmath}	

\title{Nonequilibrium Current in the One Dimensional Hubbard Model at Half-Filling}
\author{Shunsuke Kirino and Kazuo Ueda\thanks{E-mail address: kirino@issp.u-tokyo.ac.jp}}
\inst{Institute for Solid State Physics, University of Tokyo, Kashiwanoha 5-1-5, Kashiwa, Chiba 277-8581, Japan.}

\abst{
 Nonlinear transport in the one dimensional Hubbard model at half-filling
 under a finite bias voltage is investigated by the adaptive
 time-dependent density matrix renormalization group method.
 For repulsive on-site interaction, dielectric breakdown of the Mott insulating
 ground state to a current-carrying nonequilibrium steady state is
 clearly observed when the voltage exceeds the charge gap.
 It is found that by increasing the voltage further
 the current-voltage characteristics are scaled only by the charge gap
 and the scaling curve exhibits almost linear dependence on the voltage
 whose slope is suppressed by the electron correlation.
 In the case of attractive interaction
 the linear conductance is the perfect one $2e^2/h$
 which agrees with the prediction by the Luttinger liquid
 theory.
}

\kword{Hubbard model, Mott insulator, dielectric breakdown,
nonequilibrium steady state, adaptive time-dependent DMRG}

\begin{document}
\maketitle

Emergence of rich variety of different states of matters is a
consequence of electron-electron interaction.
The Hubbard model is a prototypical interacting electron system and
shows different phases depending on lattice structure, filling and
interaction.
When the band is half-filled and the Coulomb repulsion is sufficiently strong,
the charge excitations involve a finite energy gap $\Delta_c$,
and this fact is a manifestation of the Mott insulating ground state.
In one dimension (1-D) intriguing properties of the model including the
Mott transition have been clarified by various analytic approaches:
the Tomonaga-Luttinger liquid theory, the Bethe Ansatz and the conformal
field theory\cite{1DHubbard}.
Therefore concerning equilibrium properties one can say that the 1-D
Hubbard model is the best studied model in depth.

Instead of chemical doping which is commonly used to realize metal-insulator
transitions, one can also apply a bias voltage to break an
insulating phase\cite{Tokura}.
This process, dielectric breakdown of a Mott insulator, may be called as
{\it a nonequilibrium metal-insulator transition}.
However, no systematic theoretical study on the breakdown of Mott
insulators have been performed due to the difficulty to treat the
nonequilibrium states of strongly correlated systems.

Recently the adaptive time-dependent density matrix renormalization
group (TdDMRG) algorithm was developed\cite{TdDMRG}, which is an extension
of the DMRG\cite{DMRG} method to time-dependent problems.
This technique has been used as a powerful numerical approach to
nonequilibrium problems in one spatial dimension with strong
correlation, such as single quantum dot system under finite bias
voltages\cite{OurPreviousPaper} and the interacting resonant level
model\cite{SaleurSchmitteckert}.

Oka and Aoki utilized the TdDMRG method to study the breakdown of the
Mott insulating phase of the 1-D Hubbard model driven by an external
electric field\cite{Oka}.
They demonstrated that the phenomenological expression for the
transition probability which is obtained
by replacing the band gap in the Landau-Zener formula by the many-body
charge gap $\Delta_c$ is consistent with the $U$ dependence of the
threshold electric field.
They could discuss, however, only the threshold and
it is necessary to investigate current-voltage (I-V)
characteristics beyond the breakdown to elucidate nature of nonequilibrium steady
states of the 1-D Hubbard model.

In this Letter the 1-D Hubbard model with a finite bias voltage is
studied by the TdDMRG method which enables us to obtain for the first
time reliable numerical results on currents.
These results are clear manifestation that various nonequilibrium phenomena
in strongly correlated 1-D systems have become within the reach of
theoretical investigations.

We determine the I-V characteristics
for the repulsive 1-D Hubbard model at half-filling,
and show that nonzero steady current appears when the bias voltage
exceeds $\Delta_c$.
By increasing the voltage beyond $\Delta_c$
current is scaled only by the $\Delta_c$ if the voltage is
small compared to the band width and the scaling curve has almost linear
region with its slope suppressed by the correlation effect compared to a
band insulator.
Concerning the attractive case, low energy properties of the 1-D
Hubbard model at half-filling are classified into the Luther-Emery
liquids which are characterized by gapless charge excitations and gapful
spin excitations, and the ground state of the model has two degenerated
quasi long range orders, superconducting and CDW ones\cite{Solyom}.
We show that the linear conductance of the attractive Hubbard model is
precisely given by the perfect conductance $2e^2/h$.

We consider the 1-D Hubbard chain at half-filling with an applied DC
voltage.
Our main target in this Letter is the nonequilibrium steady
states of the system at $T=0$.
Although the electric potential inside the system should be determined
self-consistently, we neglect the change of the electric potential due
to the charge redistribution.
Since the TdDMRG method is applicable only to finite systems, it is important
to reduce system size dependence of the results.
Thus we take for the voltage term the simplest model in which the
potential difference is confined solely to the central bond.
In doing so the current-voltage characteristics can be addressed by the
finite system calculation as we will show below.

In order to realize nonequilibrium steady states in the numerical
calculation, we first obtain the ground state wave function of the
system without the voltage by the standard DMRG method.
Then we calculate the time evolution of the wave function after the
switching-on of the bias voltage using the TdDMRG algorithm\cite{TdDMRG}.
The nonequilibrium steady state is described by the wave
function after some transient period.
Putting the above information, the Hamiltonian is written as
\begin{align}
\label{Hamiltonian}
 H(\tau) &= H_L + H_R
 - t' \sum_{\sigma} (c_{l \sigma}^{\dagger} c_{r \sigma} + h. c.)
 \notag \\
 &+ \frac{eV}{2} \theta(\tau) \left(N_L - N_R \right), \\
 H_{\alpha} &=
 -t \sum_{i, i+1 \in \alpha, \sigma}
   (c_{i \sigma}^{\dagger} c_{i+1 \sigma} + h. c.) \notag \\
 &+ U \sum_{i \in \alpha} c_{i \uparrow}^{\dagger} c_{i
 \downarrow}^{\dagger} c_{i \downarrow} c_{i \uparrow} \,\,\,\,\,\,\,\,\,\,\,\,
 (\alpha = L, R),
\label{Hubbard_Hamiltonian}
\end{align}
where $L (R)$ represents the left (right) half of the system,
$c_{i \sigma}$ annihilates an electron with spin $\sigma$ at $i$th site,
$\tau$ is the time variable, $t$ the hopping amplitude,
$U$ the Coulomb energy, $V$ the applied voltage and
$N_{\alpha} \equiv \sum_{i \in \alpha, \sigma} c_{i \sigma}^{\dagger} c_{i \sigma}$.
$l (r)$ is the index of rightmost (leftmost) site in $L (R)$ and
$t'$ is the hopping between the $l$th and $r$th sites.
In this Letter we concentrate on $t'= t$ case for simplicity.
The bias voltage is turned on according to the smoothed step function
$\theta(\tau) \equiv (1+\exp{[(\tau_0-\tau)/\tau_1}])^{-1}$
in order to mimic adiabatic switching-on and to soften transient behaviors.
We fix $\tau_0=4\hbar/t$ and $\tau_1=\hbar/t$ throughout this Letter.

At half-filling an electron-hole transformation for one species of spin,
known as the Shiba transformation\cite{ShibaTransformation},
\begin{align}
 c_{j \uparrow} \rightarrow c_{j \uparrow},\,
 c_{j \downarrow} \rightarrow (-1)^j c_{j \downarrow}^{\dagger},
 \label{Shiba}
\end{align}
maps the $U>0$ ($U<0$) Hubbard model to the $U<0$ ($U>0$) one.
The charge and spin degrees of freedom are interchanged by this
transformation.
For example, the bias voltage term is transformed to the
Zeeman term, and the current operator is replaced with the
spin current operator, and vice versa.
This mapping is useful to interpret $U<0$ results with the knowledge
of the model with $U>0$.

\begin{figure}[!h]
 \includegraphics[width=7.8cm,height=5cm]{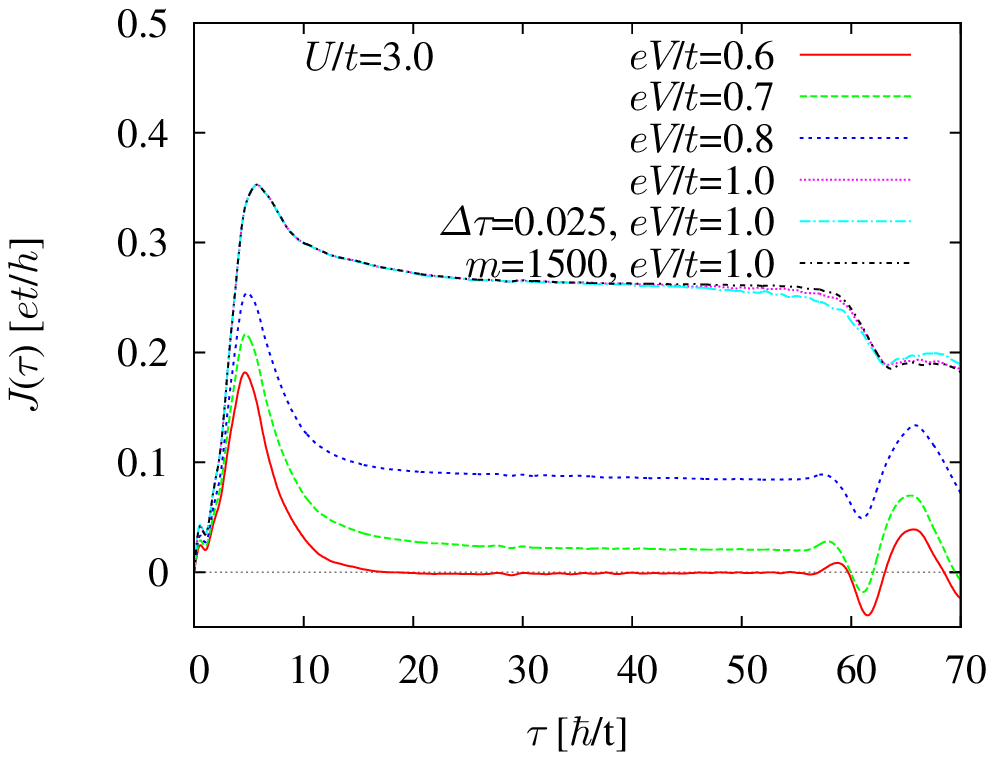}
 \includegraphics[width=8.0cm,height=5cm]{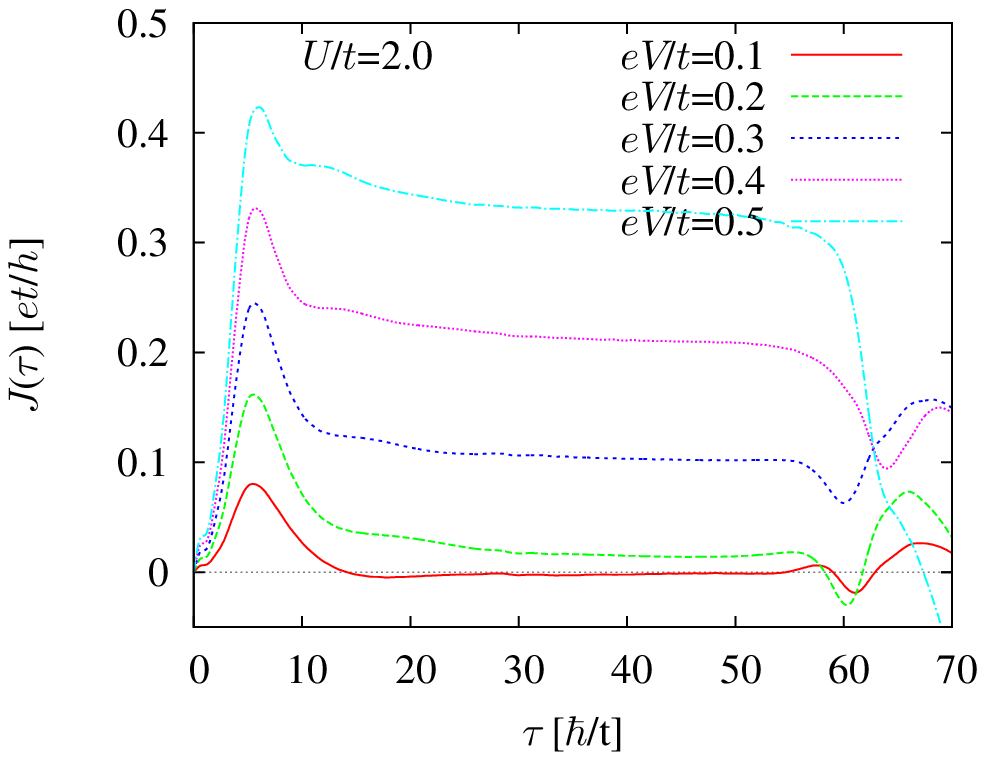}
 \includegraphics[width=7.5cm,height=5cm]{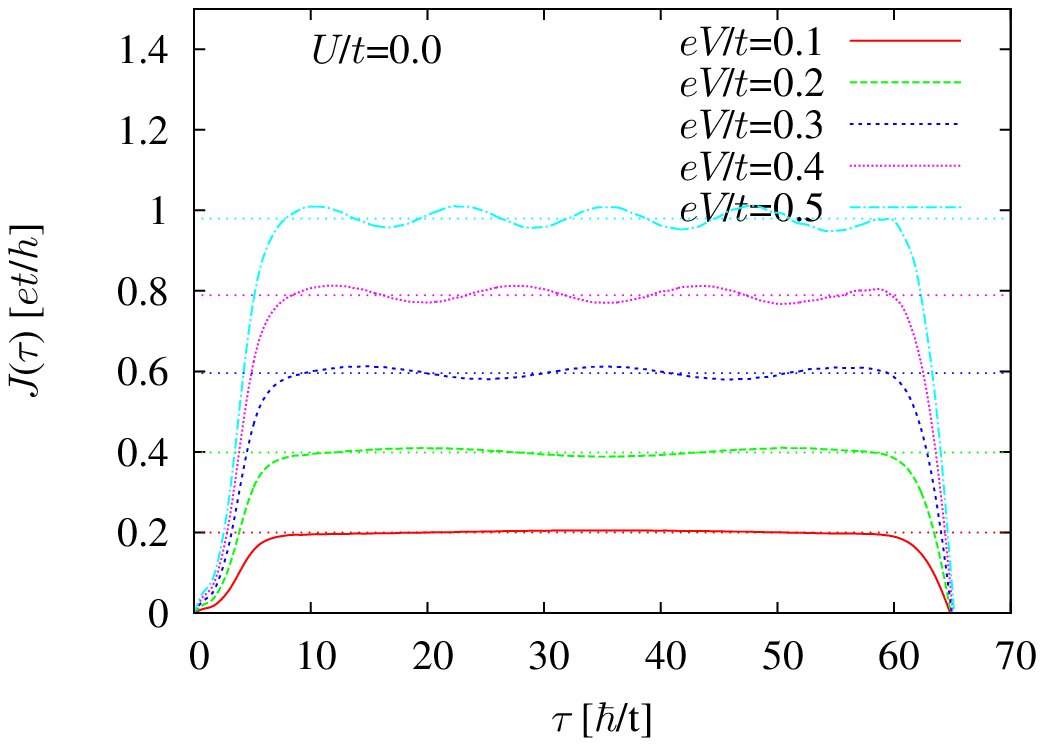}
 \includegraphics[width=7.5cm,height=5cm]{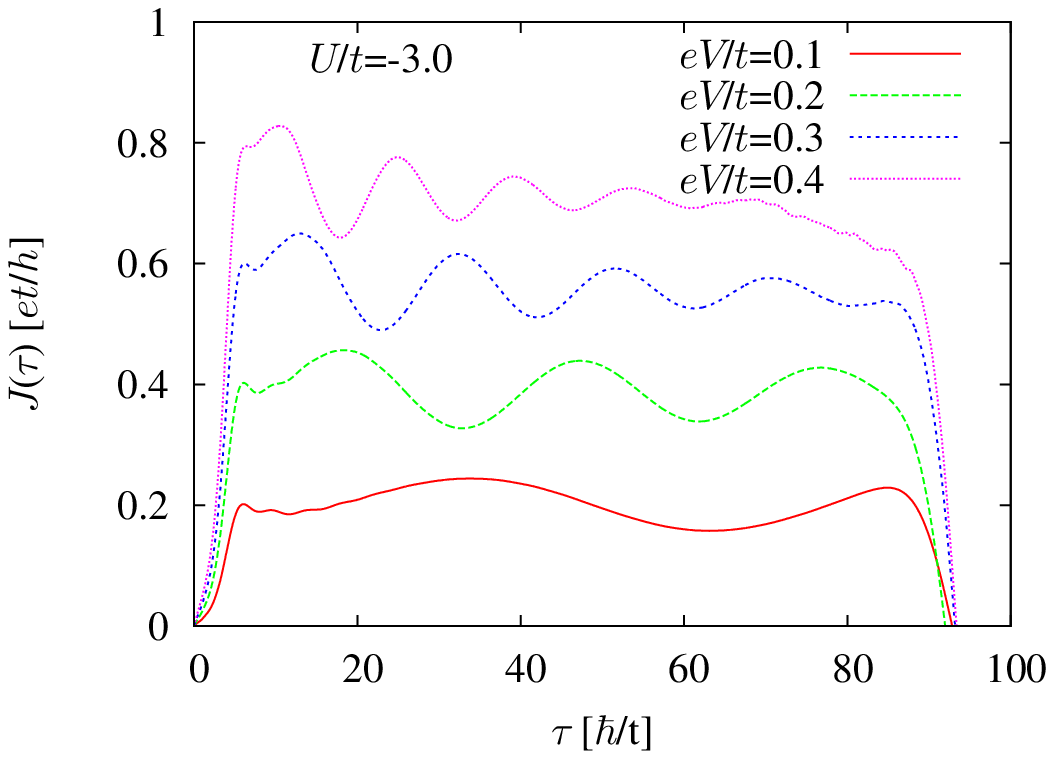}
 \caption{\label{J-t}
 (Color online)
 Time dependence of the current after the switching-on of the
 bias voltage for $L=120$, $\Delta \tau=0.05$ and $m=1200$.
 Horizontal lines in the figure for $U=0$ are the exact steady currents
 for $L=\infty$ calculated using Keldysh formalism and show that the
 steady currents can be accurately obtained by the TdDMRG calculation.
 For $U/t=3$ and $eV/t=1$ we also plot the results for
 $(\Delta \tau, m) = (0.025, 1200)$ and $(0.05, 1500)$
 to check the two types of errors, the Trotter
 error and the truncation error.
 The results are almost converged for both $\Delta \tau$ and $m$.
 }
\end{figure}

In a typical TdDMRG implementation the time evolution operator is
represented by the Suzuki-Trotter decomposition and the evolution
operator at each step is efficiently operated to the wave function
within an optimal truncated Hilbert space.
Throughout this paper the TdDMRG calculations are performed keeping
$m=1200$ states and using the 2nd order Suzuki-Trotter decomposition with
time step $\Delta \tau=0.05 \hbar / t$.
Convergence of the results in the limit $m\rightarrow\infty$ and
$\Delta\tau\rightarrow 0$ is checked for $U/t=3$ and $eV/t=1$
(see Fig.\ref{J-t}).

Current between the left and right chains is defined as
$J(\tau)= e \langle \psi(\tau) | \dot N_R | \psi(\tau) \rangle$
and its time dependence after the switching-on of the bias voltage
is shown in Fig.\ref{J-t}.
For each parameter set the current starts from $0$, exhibits a certain
transient behavior and relaxes to a steady value accompanied
by an oscillatory behavior in some cases.
After a certain time the steady-like behavior of the current is disturbed.
We hereafter call the nonequilibrium state in the time interval where
the current shows a steady-like behavior as a quasi-steady state.

Termination of the quasi-steady states is caused by the reflection of the
current at the edge of the system
\cite{Dagotto, OurPreviousPaper, SchneiderSchmitteckert} and thus a
finite size effect.
Excitations generated by applying the bias voltage propagate from left
to right with a certain velocity and are reflected at the edge,
then propagate back to the left.
Eventually those reflected excitations arrive at the center and disturb
the steady flow of the current.
Because of this effect, for an accurate determination of the steady
current one has to take long enough
system size to realize complete relaxations before the disturbances.
However at the same time this effect enables us to estimate the
velocity of the wave front of the excitations from
the time until the quasi-steady behavior ends.

From Figs.\ref{J-t}, we find that $V$ dependence of the velocity of the
wave front is small in the parameter range investigated.
In the noninteracting case, excitations of the system are described by the
single particle-hole excitations in the $-2t \cos k$ band, and the Fermi
velocity is $2$ in the units of Fig.\ref{J-t}.
Concerning the $U>0$ results we do not observe significant difference
between the velocities for $eV>\Delta_c$ and $eV<\Delta_c$.
Thus the excitations which carry the initial transient current are
almost the same for both insulating and quasi-steady states.
The velocity slightly increases from $2$ by increasing $U$ at least
for $U/t \leq 3$.
This corresponds to the fact that the velocity of the charge
excitations of the ground state at half-filling is an increasing
function of $U$ (see Table \ref{Gaps}).

On the other hand for $U<0$ we see a substantial decrease of the velocity.
As we stated before, the charge excitations of the attractive case
correspond to the spin excitations of the repulsive case.
The observed values in Fig.\ref{J-t} are $1.5$ for $U/t=-2$ (not shown
in the figure) and $1.3$ for $U/t=-3$.
These numbers qualitatively agree with the exact values of the spin
velocity listed in Table \ref{Gaps}.

The oscillation the quasi-steady states exhibit is a finite size effect
because its amplitude is proportional to $1/L$
\cite{OurPreviousPaper, SchneiderSchmitteckert}.
Note that the effect of the interaction in the whole chain strongly
influences the amplitudes of the oscillations: positive $U$ suppresses
the oscillation while negative $U$ enhances.
This fact suggests that a repulsive interaction suppresses coherent
transports in the chain while an attractive interaction promote them.
One interesting point is that the frequency of the oscillation is given
by $eV$ for both $U=0$ and $U<0$.
For $U<0$ the initial state has superconducting
correlation and thus one might expect that the AC Josephson effect may
be observed.
However the frequency turns out to be always $eV$, not $2eV$.
The AC Josephson effect is possible when the phase coherence is well
developed in each of the two subsystems which are weakly linked.
In the present system, the phase coherence is difficult to develop
because the entire system is finite and furthermore the left and right
subsystems are strongly coupled, $t' = t$.
Additionally for $U<0$ and relatively large voltage the current shows
a damped oscillation.
These effects of interaction on the oscillatory behaviors provide an
interesting subject to be investigated in the future but here we will
concentrate on the current in the quasi-steady states.

\begin{table}[t]
\begin{center}
 \begin{tabular}[t]{|c|c|c|c|}
  \hline
  $U/t$ & $\Delta_c/t$ & $v_c$ & $v_s$\\
  \hline
  $1$   & $0.005$ & $2.154$ & $1.833$\\
  \hline
  $2$   & $0.173$ & $2.300$ & $1.640$\\
  \hline
  $2.5$ & $0.371$ & $2.373$ & $1.533$\\ 
  \hline
  $3$   & $0.631$ & $2.447$ & $1.425$\\
  \hline
\end{tabular}
\end{center}
 \caption{\label{Gaps}
 The charge gap $\Delta_c$, the velocity of the charge excitations and
 the velocity of the spin excitations of the half-filled 1-D
 Hubbard model, calculated from the exact expressions\cite{exact_expressions}.
 }
\end{table}
\begin{figure}[t]
\begin{center}
 \includegraphics[width=7.5cm]{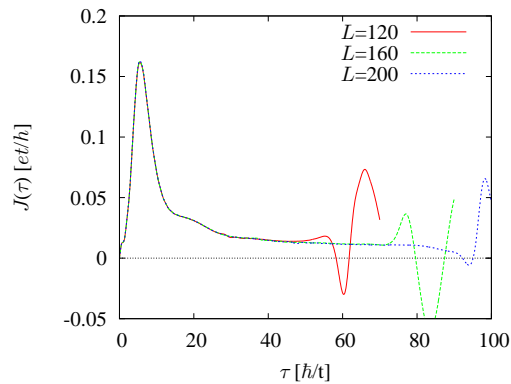}
 \caption{\label{L_dep}
 (Color online) Current as a function of time for $eV/t=0.2$,
 $U/t=2$ and various $L$, using $\Delta \tau=0.05$ and $m=1200$.
 }
\end{center}
\end{figure}

We show results of $J(\tau)$ for different system sizes in Fig.\ref{L_dep}.
No essential difference is found in $J(\tau)$ as long as the system
retains the quasi-steady behavior.
This means that the relaxation processes to real nonequilibrium steady
states can be well simulated by the finite size calculations and that
the currents of the steady states are obtained from the flat region.
In practice the steady currents are estimated from the results of
$J(\tau)$ for $L=120$ inside an interval $[30, 50]$ by the following
processes: for $U>0$ taking average, and for
$U<0$ fitting data points to the damped oscillation function
$J(\tau) \sim J(V) + \Delta J e^{-\tau/\tau_{\mathrm{damp}}}\sin(eV\tau + \theta)$.
The steady currents for $U>0$ obtained in this way are slightly
overestimated because the relaxation processes are not completely
finished in the interval, but its qualitative behavior is not influenced
by this treatment.

In order to compare I-V characteristics of the Mott insulator
with those of band insulator, we also calculate nonlinear currents in a
(noninteracting) band insulator obtained
by replacing the on-site interaction term in eq.\eqref{Hubbard_Hamiltonian}
by an alternating potential $(\Delta_b/2) \sum_{j \sigma}(-1)^{j}n_{j \sigma}$.
This term modifies the dispersion relation from $\epsilon_k = -2t\cos k$
to $\pm \sqrt{\epsilon_k^2 + (\Delta_b/2)^2}$ and opens a band gap $\Delta_b$.

\begin{figure}[t]
\begin{center}
 \includegraphics[width=5.5cm,angle=-90]{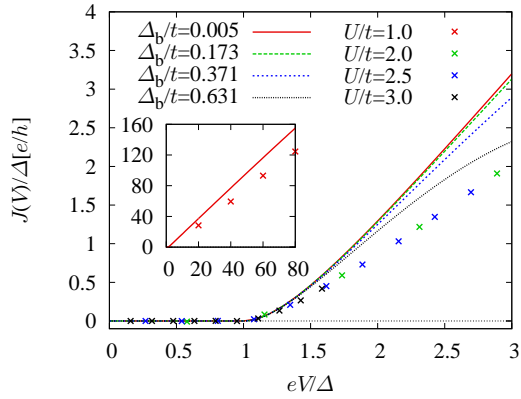}
 \caption{\label{J-V}
 (Color online)
 I-V characteristics of the Mott insulator (the 1-D Hubbard
 model) and the band insulator (see text for explanation of the model)
 scaled by the energy gap $\Delta$.
 The results of the band insulator are exactly calculated using Keldysh
 formalism and the limit $L\rightarrow \infty$ is taken.
 The results of the Mott insulator are obtained via the TdDMRG
 calculation for $L=120$.
 Values of the band gap $\Delta_b$ are chosen to be the same as
 the charge gap $\Delta_c$.
 (inset) The same I-V characteristics in a larger scale.
 }
\end{center}
\end{figure}

In Fig.\ref{J-V} we show the I-V characteristics scaled by
the gap $\Delta$, which is $\Delta_c$ for the Mott insulator and
$\Delta_b$ for the band insulator.
By the definition of the energy gap, when $eV>\Delta$ the charge
excitations are allowed, resulting in a finite current.
This behavior, dielectric breakdown, is beautifully reproduced by our results.
Furthermore, in each case $J(V)$ for different values of $\Delta$ form a single
curve when the voltage is small compared with the band width $D \simeq 4t$.
When $eV$ becomes comparable to $D$ the energy
dependence of the density of states of the band is not negligible,
and therefore $J(V)$ deviates from the scaling curve.
For $\Delta \ll eV \ll D$ the scaling curve shows
almost linear dependence, as shown in the inset of Fig.\ref{J-V}.
For the band insulator the slope is $2e^2/h$, which is the value of the perfect
conductance.
On the contrary for the Mott insulator the slope is suppressed by the
electron correlation and is $1.6e^2/h$.
This surprising scaling behavior of the 1-D Hubbard model is possible
because $\Delta_c$ is exponentially small so that all current carrying
states are under the influence of strong correlation, when $eV \ll U$.

\begin{figure}[t]
\begin{center}
 \includegraphics[width=7.5cm]{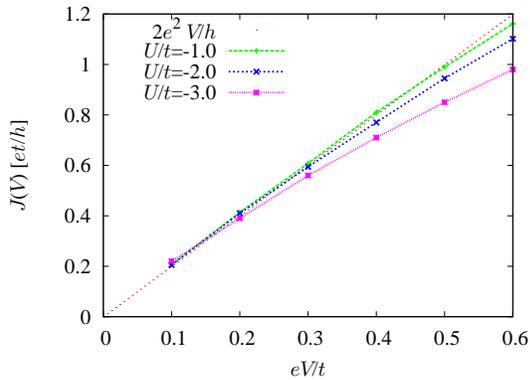}
 \caption{\label{J-V_attractive}
 (Color online) I-V characteristics of the attractive
 Hubbard model.
 }
\end{center}
\end{figure}

We show numerically obtained I-V characteristics for $U<0$
in Fig.\ref{J-V_attractive}.
Our results clearly indicate that the linear conductance for any $U<0$ is
always the same as the perfect conductance $2e^2/h$.
The gapless charge degrees of freedom of the Luther-Emery liquid are
described by the usual Luttinger liquid theory and
the conductance of the Luttinger liquid is renormalized by the
correlation exponent $K_\rho$ as $G = (2e^2/h) K_{\rho}$
\cite{KaneFisher}.
When we use the Shiba transformation eq.\eqref{Shiba},
$K_\rho$ in the expression is actually given by the correlation exponent
for the spin channel $K_\sigma$ of the repulsive model.
Then if we take the linear limit $V\rightarrow 0$ the model restore the
spin SU(2) symmetry and this ensures $K_\sigma=1$.
Accordingly, the linear conductance should be equal to $2e^2/h$ and this
agrees precisely with our results.
Away from the linear response regime $J(V)$ deviates from $2e^2V/h$
by increasing the voltage and the deviation becomes larger with
increasing $|U|$.

In summary, we have studied nonequilibrium transport phenomena in the
1-D Hubbard model at half-filling with a finite bias voltage using the
TdDMRG technique.
We have determined the I-V characteristics and found that
the current for $U>0$ in the region $eV \ll \min(D, U)$ shows a
universal behavior while the linear conductance for $U<0$ is the perfect
conductance $2e^2/h$.
These truly nonequilibrium properties were numerically addressed with
sufficient accuracy for the first time.
We believe that these reliable data about the nonequilibrium transport
of the 1-D Hubbard model provide basis for future studies and stimulate,
in particular, analytic approaches.

The authors would like to acknowledge
T. Fujii, M. Oshikawa, M. Sigrist, H. Tsunetsugu and N. Kawakami
for helpful discussions.
S. K. is supported by the Japan Society for the Promotion of Science.
This work was supported by JSPS Grant-in-Aid for JSPS Fellows
21$\cdot$6752, by Grant-in-Aid on Innovative Areas ``Heavy Electrons''
(No. 20102008) and also by Scientific Research (C) (No. 20540347).


\begin{thebibliography}{00}
 \bibitem{1DHubbard}
	 See for example, F. H. L. Essler, H. Frahm, F. G\"ohmann,
	 A. Kl\"umper and V. E. Korepin:
	 {\it The One-Dimensional Hubbard Model} (Cambridge, 2005).
 \bibitem{Tokura}
	 A. Asamitsu, Y. Tomioka, H. Kuwahara and Y. Tokura:
	 Nature (London) \mbox{\boldmath $388$} (1997) 50;
	 Y. Taguchi, T. Matsumoto and Y. Tokura:
	 Phys. Rev. B \mbox{\boldmath$62$} (2000) 7015.
 \bibitem{TdDMRG}
	 A. J. Daley, C. Kollath, U. Schollw\"ock and G. Vidal:
	 J. Stat. Mech. Theor. Exp. P04005 (2004);
	 S. R. White and A. E. Feiguin:
	 Phys. Rev. Lett. \mbox{\boldmath $93$} (2004) 076401.
 \bibitem{DMRG}
	 S. R. White: Phys. Rev. Lett. \mbox{\boldmath $69$} (1992) 2863;
	 S. R. White: Phys. Rev. B \mbox{\boldmath $48$} (1993) 10345.
	 For reviews on DMRG method see,
	 U. Scholl\"ock: Rev. Mod. Phys. \mbox{\boldmath $77$} (2005) 259;
	 K.A. Hallberg: Adv. Phys. \mbox{\boldmath $55$} (2006) 477.
 \bibitem{OurPreviousPaper}
	 S. Kirino, J. Zhao, T. Fujii and K. Ueda:
	 J. Phys. Soc. Jpn. \mbox{\boldmath $77$} (2008) 084704.
 \bibitem{SaleurSchmitteckert}
	 E. Boulat, H. Saleur and P. Schmitteckert:
	 Phys. Rev. Lett. \mbox{\boldmath $101$} (2008) 140601.
 \bibitem{Oka}
	 T. Oka and H. Aoki: Phys. Rev. Lett. \mbox{\boldmath $95$} (2005) 137601;
	 T. Oka and H. Aoki:
	 {\it Lecture Notes in Physics} \mbox{\boldmath $762$}
	 (Springer, 2009).
 \bibitem{Solyom}
	 J. S\'{o}lyom: Adv. Phys. \mbox{\boldmath $28$} (1979) 201.
 \bibitem{ShibaTransformation}
	 H. Shiba: Prog. Theor. Phys. \mbox{\boldmath $48$} (1972) 2171.
 \bibitem{SchneiderSchmitteckert}
	 G. Schneider and P. Schmitteckert:
	 cond-mat 0601389 (2006).
 \bibitem{Dagotto}
	 K. A. Al-Hassanieh, A. E. Feiguin, J. A. Riera, C. A. Busser
	 and E. Dagotto:
	 Phys. Rev. B \mbox{\boldmath $73$} (2006) 195304.
 \bibitem{exact_expressions}
	 The exact expressions for the charge gap $\Delta_c$, the charge
	 velocity $v_c$ and the spin velocity $v_s$ are given in eqs.(6.35),
	 (7.26) and (7.21) in ref.\cite{1DHubbard}, respectively.
 \bibitem{KaneFisher}
	 W. Apel and T. M. Rice: Phys. Rev. B \mbox{\boldmath $26$} (1982) 7063;
	 C. L. Kane and M. P. A. Fisher:
	 Phys. Rev. Lett. \mbox{\boldmath $68$} (1992) 1220.
\end{thebibliography}
\end{document}